\pdfoutput=1
\documentclass[11pt,twoside,a4paper,cmspaper,final,collab]{cms-tdr}
\def\svnVersion{e308dcd}\def\svnDate{2024/02/26}\def\cmsCernNoTag{CERN-EP-2023-094}\def\cmsCernDate{\today}\def\cmsMessage{Published in Physical Review C as \href{http://dx.doi.org/10.1103/PhysRevC.109.024914}{\doi{10.1103/PhysRevC.109.024914}.}}
\begin{document}\cmsNoteHeader{HIN-21-011}

\newcommand{\KT}{\ensuremath{K_{\mathrm{T}}}\xspace}
\newcommand{\PbPb}{\ensuremath{\text{PbPb}}\xspace}
\newcommand{\Aq}{\ensuremath{A(q)}\xspace}
\newcommand{\Bq}{\ensuremath{B(q)}\xspace}
\newcommand{\Cq}{\ensuremath{C_2(q)}\xspace}
\newcommand{\DRq}{\ensuremath{DR(q)}\xspace}
\newcommand{\NpartAv}{\ensuremath{\langle N_{\text{part}}\rangle}\xspace}
\newcommand{\alphaAv}{\ensuremath{\langle \alpha \rangle}\xspace}
\newlength\cmsTabSkip\setlength{\cmsTabSkip}{1ex}
\newcommand{\cOverfm}{\ensuremath{\unit{$c$/fm}}\xspace}
\providecommand{\cmsTable}[1]{\resizebox{\textwidth}{!}{#1}}

\cmsNoteHeader{HIN-21-011}
\title{Two-particle Bose--Einstein correlations and their L\'evy parameters in \PbPb collisions at \texorpdfstring{$\sqrtsNN = 5.02\TeV$}{sqrt(sNN) = 5.02 TeV}}

\date{\today}

\abstract{
Two-particle Bose--Einstein momentum correlation functions are studied for charged-hadron pairs in lead-lead collisions at a center-of-mass energy per nucleon pair  of $\sqrtsNN = 5.02\TeV$.  The data sample, containing 4.27\ten{9} minimum bias events corresponding to an integrated luminosity of 0.607\nbinv, was collected by the CMS experiment in 2018.  The experimental results are discussed in terms of a L\'evy-type source distribution.  The parameters of this distribution are extracted as functions  of particle pair average transverse mass and collision centrality.  These parameters include the L\'evy index or shape parameter ($\alpha$), the L\'evy scale parameter ($R$), and the correlation strength parameter ($\lambda$).  The source shape, characterized by $\alpha$,  is found to be neither Cauchy nor Gaussian, implying the need for a full L\'evy analysis. Similarly to what was previously found for systems characterized by Gaussian source radii, a hydrodynamical scaling is observed for the L\'evy $R$ parameter. The $\lambda$ parameter is studied in terms of the core-halo model.
}

\hypersetup{%
pdfauthor={CMS Collaboration},%
pdftitle={Measurement and analysis of two-particle HBT correlations and their Levy-stability parameters in sqrt(sNN)=5.02 TeV PbPb collisions},%
pdfsubject={CMS},%
pdfkeywords={CMS, heavy ion collision, momentum correlation, Levy HBT}} %

\maketitle

\section{Introduction}

The effect of quantum statistics is evident in the momentum correlations observed for identical particles emitted in high energy hadronic and nuclear collisions. These correlations were first studied using pion pairs produced in proton-antiproton collisions~\cite{Goldhaber:1959mj}, where they were explained by the Bose--Einstein nature of pions~\cite{Goldhaber:1960sf}. Subsequent to the first Bose--Einstein momentum correlation measurements, it was shown~\cite{Glauber:1962tt,Glauber:2006zz,Glauber:2006gd} that the particle correlations are closely related to earlier intensity correlation measurements used to determine the angular diameters of stars, as performed by R. Hanbury Brown and R.Q. Twiss~\cite{HanburyBrown:1956bqd}, and are now referred to as HBT correlations. The observed momentum correlations can be related to the spatio-temporal particle emission probability, the so-called ``source distribution", through a Fourier transform. This allows for a determination of the spatial structure of the particle-emitting source at the femtometer scale.

Measuring and interpreting such momentum correlations is called ``femtoscopy" and these studies have significantly advanced our understanding of the processes governing heavy ion collisions~\cite{Lednicky:2001qv}.
The fluid nature of the strongly coupled quark-gluon plasma created in heavy ion collisions was established, in part, 
by Bose--Einstein correlation measurements~\cite{Adler:2004rq}. 
In particular, the pair transverse momentum dependence of the Gaussian source radii~\cite{NA44:2001hbu,Adler:2004rq,Afanasiev:2009ii} can be explained as a consequence of the hydrodynamical expansion of the medium~\cite{Makhlin:1987gm,Csorgo:1995bi}.
In addition, the measured source radii provide information on the transition from a quark-gluon plasma to a hadronic state of matter~\cite{Csanad:2004mm,Pratt:2008qv} and have also been important for extending our understanding of quantum chromodynamics to new regions of phase space~\cite{Lacey:2014wqa}.

In past femtoscopic measurements, either Gaussian~\cite{Adler:2004rq,STAR:2004qya,ALICE:2015hvw,ALICE:2015tra} or Cauchy~\cite{CMS:2017mdg,ATLAS:2017shk} source distributions
have generally been assumed. 
However, recent high precision correlation measurements  in gold-gold (AuAu) collisions at a center-of-mass energy per nucleon pair of $\sqrtsNN = 200\GeV$ at BNL RHIC~\cite{PHENIX:2017ino} and in beryllium-beryllium collisions at $150A\GeVc$ beam momentum at CERN SPS~\cite{NA61SHINE:2023qzr} have
shown that neither the Gaussian nor the Cauchy approximation can adequately reproduce the measured results. Rather, a
generalization of these distributions, the so-called L\'evy alpha-stable
distribution~\cite{UchaikinZolotarev+2011}, is required to describe the data. The L\'evy exponent $\alpha$ quantifies the deviation of the source shape from a Gaussian distribution and may be influenced by various physical processes~\cite{Csorgo:2004sr,Csorgo:2005it,Csanad:2007fr,Kincses:2022eqq,Korodi:2022ohn}.  

In this paper, the source geometry is investigated for lead-lead (\PbPb) collisions at $\sqrtsNN = 5.02\TeV$ by measuring quantum-statistical two-particle Bose--Einstein correlations of unidentified charged hadrons. The measured correlation functions are fitted with a formula based on a L\'evy source distribution, allowing for the possibility of a non-Gaussian source shape. The extracted source parameters are studied as functions of pair transverse momentum and collision geometry.

The paper is structured as follows: The theory of Bose--Einstein correlations and the connection between the two-particle correlation function and the geometry of the particle emitting source are discussed in Section~\ref{sec:BEcorr}. The experimental setup is described in Section~\ref{sec:detector}. The analysis details and systematic uncertainties are discussed in Section~\ref{sec:measurement}. The results are presented and discussed in Section~\ref{sec:results}, with a summary presented in Section~\ref{sec:summary}. Tabulated results are provided in the HEPData record for this analysis~\cite{hepdata}.

\section{Bose--Einstein correlations} 
\label{sec:BEcorr}

The general definition of the two-particle momentum correlation function is given by
\begin{linenomath}
\begin{equation}\label{e:c2def}
C_2(p_1,p_2)=\frac{N_{12}(p_1,p_2)}{N_1(p_1)N_2(p_2)},
\end{equation}
\end{linenomath}
where $p_1$ and $p_2$ are the particle four-momenta, $N_1(p_1)$ and $N_2(p_2)$ are the single particle momentum distributions, 
and $N_{12}(p_1,p_2)$ is the two-particle momentum distribution. 
For identical bosons in high-multiplicity heavy ion collisions, the main source of correlations is the Bose--Einstein effect~\cite{PHENIX:2017ino}. 
This can be understood if one expresses the momentum distributions in terms of the source distribution, $S(x,p)$, which is
defined as the probability density distribution of creating a particle with
four-momentum $p$ at spacetime point $x$.  
As shown in Refs.~\cite{Csorgo:1999sj,Wiedemann:1999qn}, in the approximation that $p_1\approx p_2\approx K$, with $K=(p_1+p_2)/2$ being the average four-momentum, the relationship between the correlation function and the source distribution becomes  
\begin{linenomath}
\begin{equation}\label{e:C2expr}
C^{(0)}_2(Q,K)=1+\frac{\abs{\widetilde S(Q,K)}^2}{\abs{\widetilde S(0,K)}^2} ,
\end{equation}
\end{linenomath}
where $\widetilde S(Q,K)$ is the Fourier transform of the source distribution,
\begin{linenomath}
\begin{align}\label{e:tildeSdef}
\widetilde S(Q,K)=\int S(x,K) \re^{iQx} \rd^4x ,
\end{align}
\end{linenomath}
with $Q=p_1-p_2$ denoting the relative four-momentum of the two particles.  In Eq.~(\ref{e:C2expr}), the superscript ${(0)}$ denotes that final-state interactions of hadrons, such as the Coulomb interaction, are ignored. The effect of the Coulomb
interaction will be discussed later in this section. 

The above expressions imply $C^{(0)}_2(Q=0,K) = 2$. This cannot be directly verified since the two-particle resolution limits the measurement of $C_2$ at small values of relative momentum $Q$. However, based on extrapolations of observed correlation functions, it is found that, in general, $C^{(0)}_2(Q \to 0) < 2$. This observation has led to the development of a ``core-halo" picture~\cite{Bolz:1992hc,Csorgo:1994in,Kisiel:2006is,ALICE:2020ibs,Mihaylov:2023pyl},
where the particle emitting source is divided into two parts, a core of primordial hadrons (including also the decay products of short lived resonances) and a halo of long-lived resonances. The latter is experimentally unresolvable due to its large size that corresponds to very small momentum difference in Fourier space.
Taking $S$ to represent the core part of the source and $\lambda$ as the square of the core fraction, one obtains
\begin{linenomath}
\begin{align}\label{e:C2:corehalo}
C^{(0)}_2(Q,K)&=1+\lambda \frac{\abs{\widetilde S(Q,K)}^2}{\abs{\widetilde S(0,K)}^2},
\end{align}
\end{linenomath}
with
\begin{linenomath}
\begin{align}
\lambda = \left(\frac{N_{\text{core}}}{N_{\text{core}}+N_{\text{halo}}}\right)^2.
\label{e:lambdadef}
\end{align}
\end{linenomath}
Here $N_{\text{core}}$ is the number of hadrons in the core (for a given species) and $N_{\text{halo}}$ is the number of hadrons in the halo. The $\lambda$ parameter is called the correlation strength, because it qualitatively describes how strong the two-particle correlation is. Other approaches to account for the effect of resonances are discussed in Refs.~\cite{Kisiel:2006is,ALICE:2020ibs,Mihaylov:2023pyl}.

Using Eq.~(\ref{e:C2:corehalo}), the correlation function can be calculated for an assumed source distribution, with the parameters of the source distribution then determined by fits to the experimental results. The Gaussian, Cauchy, and a generalization of these two distribution functions, the so-called L\'evy alpha-stable distribution, have been studied in Ref.~\cite{PHENIX:2017ino}. The motivation behind these studies is that the mean free path increases over time in an expanding hadron gas. Hence, if one describes the evolution of hadrons by generating random steps, the distribution of the size of these steps may not have a finite variance. In this case, the central limit theorem cannot be applied, thus the limiting distribution of the sum of step sizes for a given hadron is not a Gaussian. This effect has been referred to as ``anomalous diffusion.'' However, a more generalized version of the central limit theorem is applicable and the limiting distribution in this case falls in the class of symmetric L\'evy alpha-stable distributions~\cite{Metzler:1999zz,Csorgo:2003uv}, defined as
\begin{linenomath}
\begin{align}\label{e:Levydef}
\mathcal{L}(\boldsymbol{r};\alpha,R)=\frac{1}{(2\pi)^3}\int \rd^3\boldsymbol{q}\, \re^{i\boldsymbol{q}\cdot\boldsymbol{r}} \re^{-\frac{1}{2}\abs{\boldsymbol{q}R}^{\alpha}},
\end{align}
\end{linenomath}
with $\boldsymbol{r}$ being the variable of the distribution and $\boldsymbol{q}$ being an arbitrary integration variable. In this paper, $\mathcal{L}(\boldsymbol{r};\alpha,R)$ is assumed for the shape of the source distribution, \ie, $S$ mentioned above (and used in Eq.~(\ref{e:tildeSdef})) is assumed to be a L\'evy distribution $\mathcal{L}$, with parameters $\alpha$ and $R$ depending on average momentum $K$. In this case, $R$ is the L\'evy scale parameter and $\alpha$ is the L\'evy index of stability. Stability means 
that for $\alpha\leq 2$ this distribution represents the limit of properly normalized sums of independent random variables, 
and the sum of L\'evy distributed random variables (with the same stability index) is also L\'evy distributed. 
The Gaussian distribution is obtained with $\alpha = 2$ and the Cauchy distribution is obtained with  $\alpha = 1$. A value of $\alpha>2$ is not physically allowed for a stable L\'evy distribution (as discussed in Ref.~\cite{UchaikinZolotarev+2011}). The $\alpha$ parameter describes the shape of the source and its deviation from the Gaussian distribution. In case of a Gaussian source, $R$ is simply the quadratic mean of the distribution. However, for $\alpha<2$, the stable distributions do not have a second moment. In those cases, $R$ is still proportional to the full width at half maximum (with the proportionality constant depending on $\alpha$), and therefore represents the spatial scale of the source.
The above considerations lead to the following correlation function for a general value of $\alpha$~\cite{Csorgo:2003uv}:
\begin{linenomath}
\begin{align}\label{e:Levy:FT}
C^{(0)}_2(q)=1+\lambda \re^{-(qR)^{\alpha}}.
\end{align}
\end{linenomath}
Here $q$ is a one-dimensional (1-D) variable, the magnitude of the spatial part of $Q$ in a given frame: $q=\abs{\boldsymbol{Q}}$. 
The dependence on the average momentum $K$ is carried by the parameters: $\lambda$, $\alpha$, and $R$~\cite{PHENIX:2017ino}. 

The above formulas are based on a plane wave analysis and are only valid for interaction-free particles. Final-state interactions distort this simple picture. 
In the so-called Bowler--Sinyukov method~\cite{Sinyukov:1998fc}, the most significant final-state interaction in case of charged particles, the Coulomb interaction, is handled by including a 
correction term $K_{C}(q;R,\alpha)$, with
\begin{linenomath}
\begin{equation}\label{e:C2:Coulomb}
C_2(q)= 1-\lambda  + \lambda  (1+\re^{-(qR)^{\alpha}})K_{C}(q;R,\alpha).
\end{equation}
\end{linenomath}
This correction can be calculated based on the Coulomb wave function, as determined by solving the two-particle Schr\"odinger equation~\cite{PHENIX:2017ino,Csanad:2019lkp}. 
Such a calculation requires numerical integration of the L\'evy distribution multiplied by the square of the absolute value of the Coulomb wave function. A parametrization of the 
results of this calculation, 
as a function of $R$ and $\alpha$, is given in Ref.~\cite{Csanad:2019lkp}. The measured correlation functions are fitted with the formula described in Eq.~(\ref{e:C2:Coulomb}).

\section{The CMS detector}
\label{sec:detector}

The CMS detector is built around a superconducting solenoid of 6\unit{m} internal diameter. Within the solenoid volume are a silicon pixel and strip tracker, 
a lead tungstate crystal electromagnetic calorimeter ($\abs{\eta}<3$), and a brass and scintillator hadron calorimeter ($\abs{\eta}<3$), 
each composed of a barrel and two endcap sections, where $\eta$ is the pseudorapidity. In addition to the barrel and endcap detectors, 
quartz-fiber Cherenkov hadron forward (HF) calorimeters ($3<\abs{\eta}<5$) complement the coverage provided by the barrel and endcap detectors 
on either side of the interaction point. These HF calorimeters are azimuthally subdivided in $\phi$ into 20\de modular wedges and further 
segmented to form $0.175{\times}0.175$ $(\Delta \eta{\times}\Delta \phi)$ ``towers'', where $\phi$ is the azimuthal angle. A muon system located outside the solenoid and 
embedded in the steel flux-return yoke is used for the reconstruction and identification of muons with $\abs{\eta} < 2.4$. 

The silicon tracker measures the properties of charged particles with $\abs{\eta} < 2.5$. During the 2018 LHC running period corresponding to the 
data used in this paper, the silicon tracker consisted of 1856 silicon pixel and 15\,148 silicon strip detector modules. 
Details on the pixel detector can be found in Ref.~\cite{Phase1Pixel}. For nonisolated particles with a transverse momentum of $1 < \pt < 10\GeVc$ and $\abs{\eta} < 2.5$,
the track resolutions are typically 1.5\% in \pt and 20--75\mum in the transverse impact parameter ($d_{\mathrm{xy}}$)~\cite{DP-2020-049}.  
Events of interest are selected using a two-tiered trigger system~\cite{CMS:2016ngn}. The first level, composed of custom hardware processors, 
uses information from the calorimeters and muon detectors to select events at a rate of around 100\unit{kHz}.
The second level, known as the high-level trigger, consists of a farm of processors running a version of the full event reconstruction 
software optimized for fast processing, and reduces the event rate to around 1\unit{kHz} before data storage. 
A detailed description of the CMS detector, together with a definition of the coordinate system and kinematic variables, can be found in Ref.~\cite{CMS:2008xjf}.

\section{Measurement details}
\label{sec:measurement}

\subsection{Event and track selection}\label{ss:event_track}

The analysis presented in this paper used 4.27\ten{9} minimum bias triggered events, corresponding to an integrated luminosity of 0.607\nbinv~\cite{CMS:LUM-17-003,CMS:LUM-18-001}, from \PbPb collisions collected by the CMS experiment during the 2018 LHC run at $\sqrtsNN = 5.02\TeV$. The minimum
bias events were triggered by requiring signals
above thresholds in the range of $\sim$6--12\GeV energy in at least one channel of each of the HF calorimeters~\cite{CMS:2016ngn}. Further selections were applied to reject events from beam-gas interactions and nonhadronic collisions~\cite{CMS:2016xef}.
Events were also required to have precisely one interaction vertex, reconstructed based on two or more tracks, 
and within a distance of less than 15\cm from the center of the nominal interaction point along the beam axis. 
The shapes of the clusters in the pixel detector had to be compatible with those expected from particles produced at the interaction vertex location. 
After these selections, 2.65\ten{9} usable events were obtained. The event centrality was calculated from
the transverse energy deposited in both HF calorimeters, using the methodology detailed in
Ref~\cite{CMS:2011iwn}.  The centrality describes the fraction of the total inelastic hadronic cross section, and is connected to the degree of overlap of the two incident nuclei, with 0\% corresponding to the most central (total overlap) and 100\% to the least central (least overlap) collisions.

Only the standard CMS \textit{highPurity\/} tracks defined in Ref.~\cite{CMS:2014pgm} were used for the analysis. 
Furthermore, individual particles were required to have $\pt >0.5\GeVc$ with a relative uncertainty $\delta \pt<10\%$. 
To focus on mid-rapidity Bose--Einstein correlations, the analysis was restricted to tracks with $\abs{\eta} < 0.95$. 
As the correlation function parameters depend on the average pseudorapidity of
the considered pair~\cite{Csanad:2005qr}, this limited pseudorapidity interval provides a
cleaner data sample where the correlation function shape is not influenced by the $\eta$ dependence.
Tracks were required to have at least 2 hits ($N_{\text{pixel-hit}}$) in the silicon pixel detector, to have at least 11 hits ($N_{\text{hit}}$) in the strip tracking detector, 
and to have a relative $\chi^2$ of the track fit ($\chi ^2N^{-1}_{\text{dof}}N^{-1}_{\text{layer}}$) of less than 0.18, 
where $N_{\text{dof}}$ is the number of degrees of freedom during the track fitting and $N_{\text{layer}}$ is the number of tracking layers used out of the 10 barrel and 4 pixel layers. 
A reconstructed track is only considered as a candidate track from the vertex if the significance of the separation along the beam axis ($z$) between the track and the 
best vertex, $\abs{d_{\mathrm{z}}/\sigma(d_{\mathrm{z}})}$, and the significance of the track impact parameter measured in a plane transverse to the beam, $\abs{d_{\mathrm{xy}}/\sigma(d_{\mathrm{xy}})}$, are each less than 3.

In CMS, particle identification in central to mid central \PbPb collisions is hindered by the overlapping strip tracker clusters in a high track density environment. Therefore, in this measurement, no particle identification was done, and all charged-particle tracks were used. The approach presented in Section~\ref{sec:BEcorr} is only applicable for a single particle species, hence it was assumed that all detected charged particles are pions. Depending on the transverse mass and the centrality range, 60--90\% of the investigated particles are pions~\cite{ALICE:2019hno}. It was found that this assumption does not modify the physical results, apart from the $\lambda$ parameter. This is because the HBT peak in the correlation function at low $q$ is not modified by nonidentical particle pairs or by identical particle pairs of species that have been misidentified as pions and, therefore, calculated to have the wrong $q$ value. The effect of the lack of particle identification on the $\lambda$ parameter is discussed in Section~\ref{ss:lambda}.

\subsection{Event mixing and pair selection}\label{ss:mixing}

For determining the correlation functions, two distributions are formed: a signal (actual) distribution \Aq of pairs containing the Bose--Einstein correlation, and a background distribution \Bq, using the event mixing method to correct for correlations not stemming from quantum statistics or final state interactions. The signal pair distribution is formed with pairs of particles belonging to the same event.
The  $q$ variable is taken as the absolute value of the three-dimensional (3-D) momentum difference ($\abs{q}_{\mathrm{LCMS}}$) in the longitudinally comoving frame (LCMS), which is the reference system where the longitudinal component of the average pair momentum is zero. 
This distribution is affected by various effects resulting from the event  kinematics and the detector acceptance. 
To correct for these effects a background distribution is created with particle pairs taken from different events. For this purpose, a pool of events is formed based on centrality (within a width of 10\%) and collision vertex location (within 3\cm in $z$-vertex).
Then, for each data event, a mixed event was formed using the pool associated with the data event class, making sure that each particle of this mixed event belongs to a different data event. Pairs within the mixed events were used to create the above mentioned \Bq distribution. More details about the event mixing can be found in Refs.~\cite{Achard:2011zza,PHENIX:2017ino}. 
The \Aq and \Bq distributions were divided and normalized to obtain the correlation function
\begin{linenomath}
\begin{align}
C_2(q)=\frac{A(q)}{B(q)}\frac{\int_{q_1}^{q_2} B(q) \rd{q}}{\int_{q_1}^{q_2} A(q) \rd{q}}.
\end{align}
\end{linenomath}
The integrals are performed over a range $[q_1,q_2]=[4.8,6.4]\GeVc$ where the correlation function is not expected to 
exhibit quantum-statistical features, which are only present for $q\lesssim 0.2\GeVc$~\cite{PHENIX:2017ino,CMS:2019fur}, for source sizes of several femtometers. 

Tracking efficiency correction factors were used as weights when measuring pair distributions.
Each reconstructed track was weighted by the inverse of the efficiency factor, $\epsilon _{\text{trk}}(\eta,\pt, \text{centrality})$.
The efficiency weighting factor accounts for the reconstruction efficiency $\mathcal{E}_{\text{r}}$ , and the fraction
of misidentified tracks, with $f_{\text{fake}}(\eta,\pt, \text{centrality})$, and $\epsilon _{\text{trk}}=\mathcal{E}_{\text{r}}/(1-f_{\text{fake}})$.

The obtained correlation functions exhibit effects stemming from the finite segmentation of the detectors and the imperfect tracking algorithm, 
allowing for tracks to split into two reconstructed particles, or separate tracks to merge into a single reconstructed particle. 
To remove these effects from our sample, a two-dimensional (2-D) histogram of pseudorapidity difference ($\Delta \eta$) \vs azimuthal angle difference ($\Delta \phi$) was studied for the track pairs. 
A uniform distribution outside of a small $\Delta \eta$ and $\Delta \phi$ region was found, with the detector artifacts limited to this small region. The pairs were then required to satisfy the condition
\begin{linenomath}
\begin{equation}
    \left ( \frac{\abs{\Delta \eta}}{\Delta \eta _{\text{min}}} \right )^2 + \left ( \frac{\abs{\Delta \phi}}{\Delta \phi _{\text{min}}} \right )^2 >1,
\end{equation}
\end{linenomath}
with $\Delta \eta _{\text{min}}=0.014$ and $\Delta \phi _{\text{min}}=0.022$, to select pairs free of these artifacts.

While in this analysis the correlation function is determined in terms of $\abs{q}_{\mathrm{LCMS}}$, it
can also be expressed in terms of the absolute value of the invariant momentum difference, $q_{\mathrm{inv}}$~\cite{PHENIX:2017ino,Kurgyis:2020vbz}, which is the absolute value of the momentum difference in the pair center-of-mass system. 
Correlation functions depending on the transverse and longitudinal components of the 3-D momentum difference, as well as on the energy difference, were investigated. 
It was found that the correlation functions had their maximal values for small $\abs{q}_{\mathrm{LCMS}}$, rather than  small $q_{\mathrm{inv}}$ values, indicating that the most sensitive  
1-D variable of the measured correlations is $\abs{q}_{\mathrm{LCMS}}$, hence the choice  $q\equiv \abs{q}_{\mathrm{LCMS}}$ to investigate the physics parameters.

In the LCMS, the longitudinal component of the average momentum $K$ vanishes. In this case, the Fourier transform of the source distribution $\widetilde S$ 
can be expressed in terms of the transverse component $\KT$ of the average momentum. Alternatively, the transverse mass, $\mT=\sqrt{\smash[b]{m^2+\KT^2}}$, may be used, if the mass $m$ of the investigated particle species is known. 
The current analysis is done in ranges of the $\KT$ variable, with the results expressed in terms of \mT. Here, the pion mass is assumed, 
to facilitate comparisons with other experimental results and with theory.

\subsection{Correlation functions}

\begin{figure}
\centering
\includegraphics[width=0.5\textwidth]{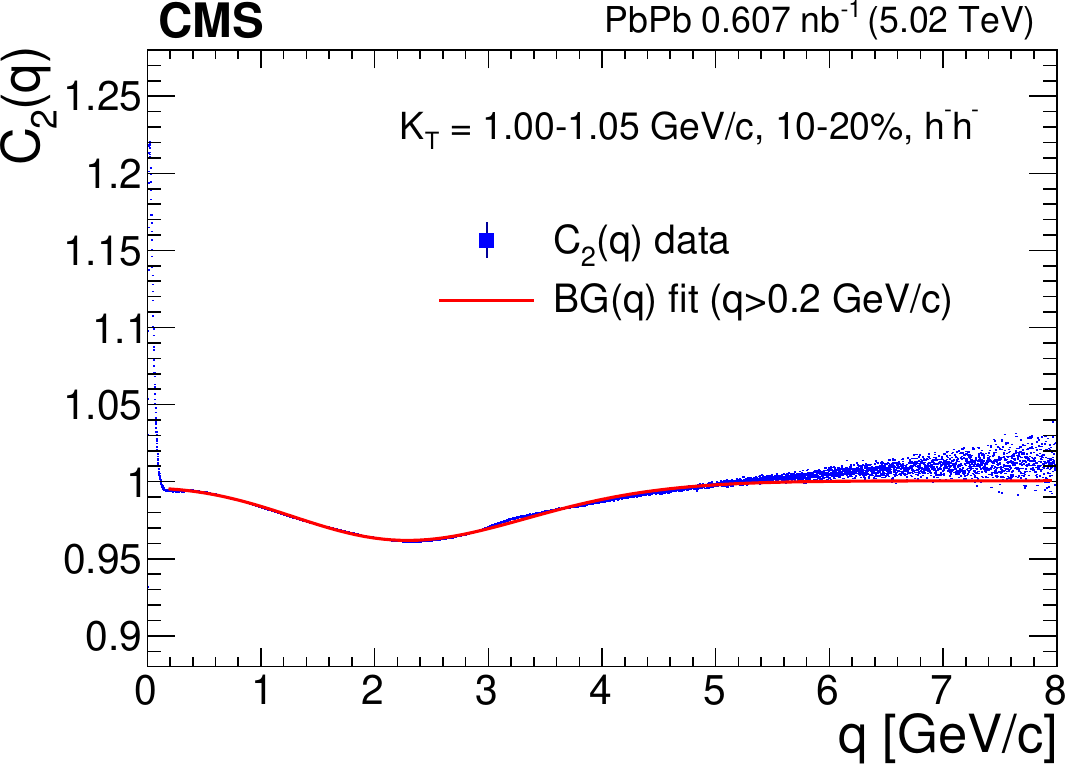}
\caption{An example of a long range background fit to the correlation function \Cq of negatively charged hadron pairs with $1.00<\KT<1.05\GeVc$ in the 10--20\% centrality bin. The Bose--Einstein peak is below $q=0.2\GeVc$, therefore the $0.2<q<8.0\GeVc$ region was used for the long range background fit.}
\label{f:Cqfit}
\end{figure}

The \Aq and \Bq distributions were measured up to $q=8\GeVc$ in 24 ranges of $\KT$ from 0.5 to 1.9\GeVc and in six centrality classes (0--5\%, 5--10\%, 10--20\%, 20--30\%, 30--40\%, 40--60\%). Negative and positive same-charged hadron pairs were measured separately. In each case, the correlation function \Cq was calculated. Although no significant difference was observed between the positively and the negatively charged pairs, the two cases were treated separately to identify minor discrepancies.
While the Bose--Einstein peak for these correlation functions occurs for $q < 200\MeVc$, a structure was also observed in the  \Cq distributions 
at higher $q$ values. This higher $q$-structure can be attributed to a number of possible sources, including energy and momentum conservation, 
resonance decays, bulk flow phenomena~\cite{CMS:2017mdg}, and minijets~\cite{CMS:2017mdg}. We addressed this by fitting the following empirically determined functional form~\cite{CMS:2019fur,CMS:2017mdg,CMS:2023jjt} to the long-range background:
\begin{linenomath}
\begin{equation}
	\label{e:bgfit}
    BG(q)=N \left( 1+\alpha_1 \re^{-(qR_1)^2} \right) \left( 1-\alpha_2 \re^{-(qR_2)^2} \right),
\end{equation}
\end{linenomath}
where $N, \alpha _1, \alpha _2, R_1, R_2$ are fit parameters. An example fit to the experimental data using Eq.~(\ref{e:bgfit}) is shown in Fig.~\ref{f:Cqfit}. The correlation function \Cq was then divided by 
the long range background function $BG(q)$ resulting in the double-ratio correlation function
\begin{linenomath}
\begin{equation}
	DR(q)=\frac{C_{2}(q)}{BG(q)}. 
\end{equation}
\end{linenomath}
The large-$q$ background is close to constant on the scale of the Bose--Einstein peak, \ie, over a range of few times 10\MeVc.  
Consequently, this correction is not as important as was found for proton-proton collisions~\cite{CMS:2019fur}, where the much smaller source radii lead to much wider correlation functions.

\subsection{L\'evy fits to correlation functions} \label{ss:fitting}

The double-ratio correlation function \DRq removes most effects other than the quantum-statistical correlation and the final state interactions. This function was fitted with  
an extension of Eq.~(\ref{e:C2:Coulomb}) that allows for a  possible residual linear background:
\begin{linenomath}
\begin{equation}
    DR(q) = N  \left[ 1-\lambda  + \lambda  (1+\re^{-(qR)^{\alpha}}) K_{C}(q;R,\alpha) \right](1+\epsilon q).
\label{drfit}
\end{equation}
\end{linenomath}
Here $R$ is the L\'evy scale parameter, $\alpha$ is the L\'evy stability index, and $\lambda$ is the correlation strength. 
The empirical normalization parameter $N$ and linear background parameter $\epsilon$ are needed
to remove any remaining background effects in the double-ratio correlation function. The Coulomb correction factor, $K_{C}(q;R,\alpha)$, 
is calculated based on Ref.~\cite{Csanad:2019lkp}.

\begin{figure}
\centering
\includegraphics[width=0.5\textwidth]{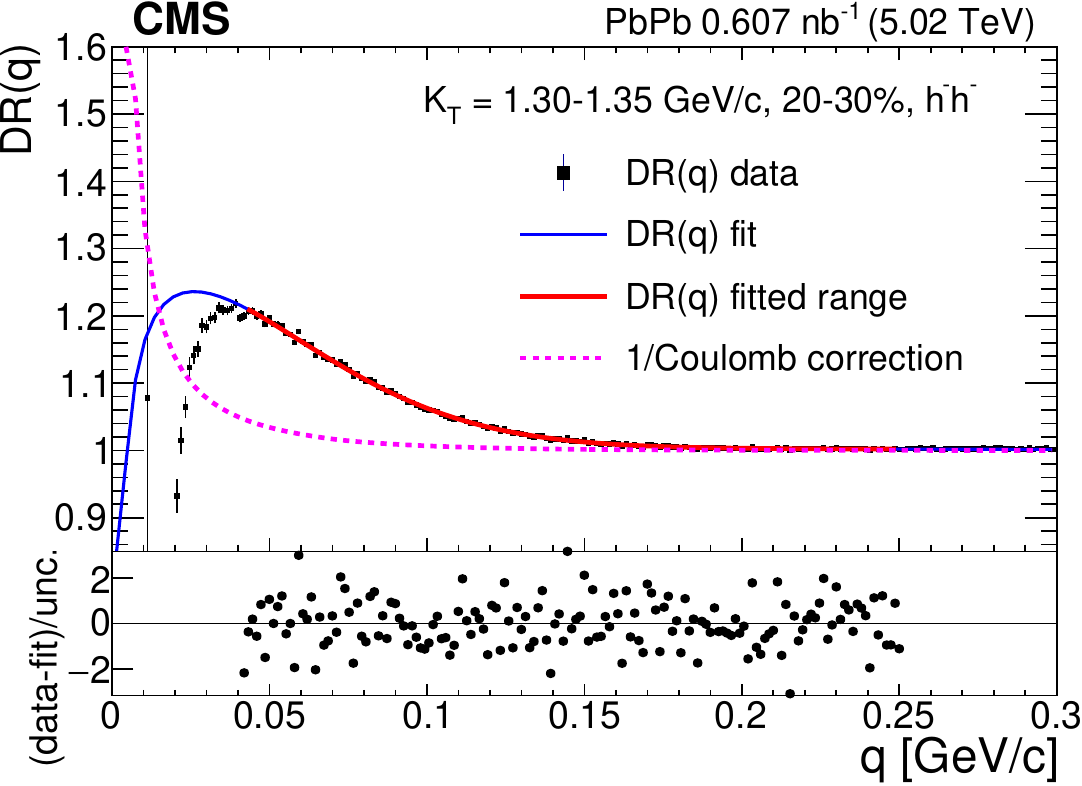}
\caption{An example fit to the double-ratio correlation function \DRq of negatively charged hadron pairs with $1.30<\KT<1.35\GeVc$ in the 20--30\% centrality bin. The error bars show the statistical uncertainties. The fitted function is shown in blue, while the red overlay indicates the  range used for the fit. The size of the Coulomb correction is indicated in magenta. The lower panel shows the deviation of the fit from the data in each bin in units of the standard deviation in that bin.}
\label{f:DRqfit}
\end{figure}

The fitting was performed using the \textsc{minuit2} package~\cite{James:1975dr,Brun:1997pa} by standard $\chi^2$ minimization. 
The asymmetric statistical uncertainties were calculated using the \textsc{minos} package~\cite{James:1975dr,Brun:1997pa}. 
A fit is deemed statistically acceptable if the confidence level calculated from the $\chi^2$ value and the number of degrees of freedom is above 0.1\%. The convergence of each fit together with a positive definite 
covariance matrix was also required. Furthermore, the stability of the fit parameters versus the binning of \DRq was tested, and no modification in the values of the parameters was found. Figure~\ref{f:DRqfit} shows the results of a typical fit. 

\begin{figure}
\centering
\includegraphics[width=0.5\textwidth]{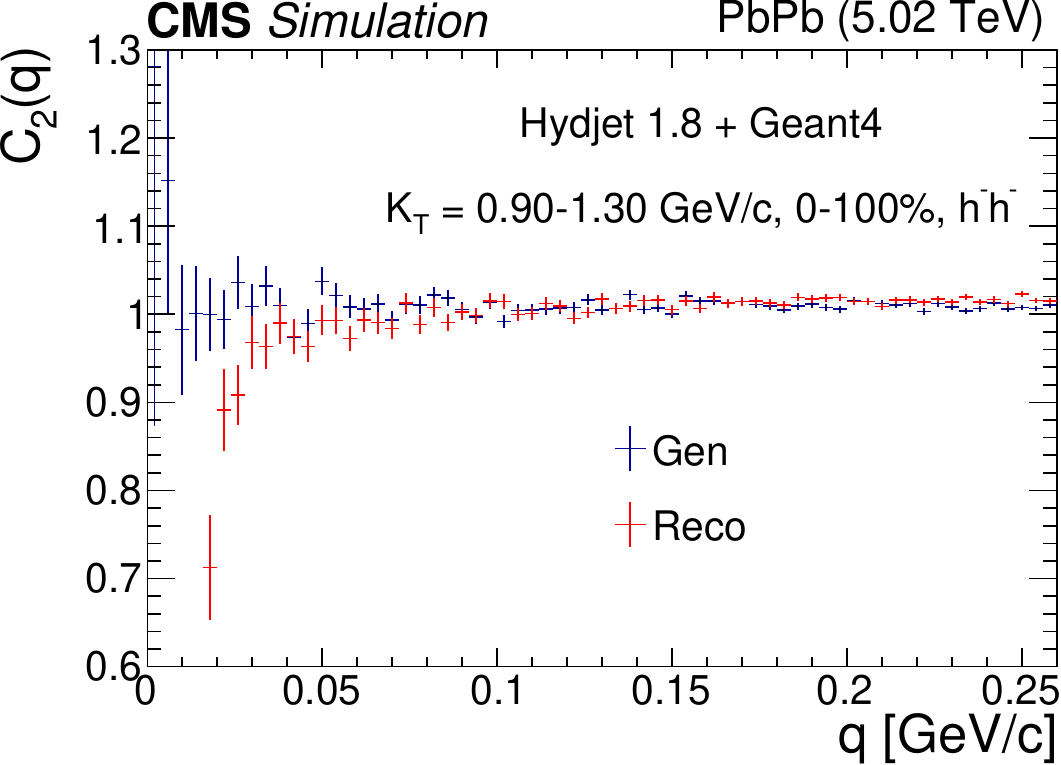}
\caption{The two-particle correlation function of negatively charged hadron pairs with $0.9<\KT<1.3\GeVc$ in the 0--100\% centrality range, calculated using Monte Carlo events with (Reco) and without (Gen) detector reconstruction effects. The error bars show the statistical uncertainties. The detector effects are most significant below approximately 50\MeVc. The quantum-statistical effects are not present in the simulations, hence there is no Bose--Einstein peak.}
\label{f:MC}
\end{figure}

Upper and lower limits in $q$ were set for the fitted region. For large $q$ values, the correlation 
no longer exhibits any features beyond those resulting from the residual long-range background.  For very small $q$ values, the data are strongly 
affected by the finite momentum resolution and pair reconstruction efficiency of the detectors, as well as the details of the pair selection. These effects were investigated with detailed Monte Carlo simulations based on \HYDJET (version 1.8, tune ``Drum''~\cite{Lokhtin:2005px}) events, where the detector response was simulated using \GEANTfour~\cite{GEANT4:2002zbu}, and a considerable decrease of pair reconstruction efficiency was found at low $q$ values, below approximately 50\MeVc, as shown in Fig.~\ref{f:MC}. Moreover, it has previously been shown in Refs.~\cite{Csanad:2019lkp,Kincses:2019rug} that the sharp decrease of the correlation function for very low $q$ values does not depend on the exact source shape, but is rather determined by the Gamow factor that corresponds to the spatial integral of the relative wave function squared multiplied by a point-like source~\cite{CMS:2019fur}. 
This very low-$q$ behavior of the fit function is found to be unaffected when calculated with various source and final-state interaction assumptions~\cite{Kincses:2019rug}. 
For the present analysis, the lower and upper fit limits were selected independently for each centrality and $\KT$ class, with the lower limit ranging between 0.024 and 0.067\GeVc and the upper limit ranging between 0.12 and 0.55\GeVc. 
The limits were selected based on the confidence level and the 
convergence of the fit and based on the property of the Bose--Einstein peak, 
which moves towards higher $q$ values for more peripheral collisions and as $\KT$ increases.

\subsection{Systematic uncertainties}\label{ss:syst}

The sources of systematic uncertainties considered in the analysis include the criteria for the event selection, the single-track selection, the pair-track selection, and the limits set on the fitted $q$ range. Furthermore, the centrality calibration is also varied to estimate the systematic uncertainty resulting from this source. For each $\KT$ and centrality class, the double-ratio correlation function \DRq was obtained and fitted with the 
nominal analysis parameters.  Variations in the fit parameters were then determined by changing, individually, each of the analysis parameters to both its lowest and highest 
plausible values, and then fitting \DRq again using these values. 
The total systematic uncertainties were obtained by adding in quadrature all of the individual contributions for positive and negative modifications in the final fit parameters separately.
Table~\ref{t:syst} lists the analysis parameters and the range through which they were varied.
These analysis parameters were separated into three categories. The ``Track and event selection'' category included the criteria on the z position of the vertex, 
the centrality edges, the selections of \pt,  $\delta \pt$, $\abs{\eta}$, $N_{\text{pixel-hit}}$, $\abs{d_{\mathrm{xy}}/\sigma (d_{\mathrm{xy}})}$, and $\abs{d_{\mathrm{z}}/\sigma (d_{\mathrm{z}})}$, respectively. 
The ``Pair selection'' category included the $(\Delta \eta ,\Delta \phi)$ pair selection. The ``Fit limits'' category included the $q_{\text{min}}$ lower, and the $q_{\text{max}}$ upper fit limits.
The asymmetric systematic uncertainties for each of the fit parameters are summarized in Table~\ref{t:Rsyst}.  Uncertainties for different $\KT$ ranges and the two charge signs are averaged 
in each centrality range. For every fit parameter, the dominant systematic uncertainty results from the modification of the fit limits. The uncertainty corresponding to the pair selection criteria is highly asymmetric for the three most central cases, it is zero in one of the directions for all three parameters.

\begin{table*}[h!]
\centering
\topcaption{The sources of the systematic uncertainties with their values in the default, lower, and upper settings. The meaning of the analysis parameters are given in Sections~\ref{ss:event_track} and~\ref{ss:mixing}.}
\begin{scotch}{cccc}
Systematic source & Default & Low & High \\ \hline
vertex z selection & ${<}15\cm$ & ${<}12\cm$ & ${<}18\cm$ \\ 
$\pt$ selection & ${>}0.5\GeVc$ & ${>}0.55\GeVc$ & ${>}0.5\GeVc$ \\ 
$\delta \pt$ selection & ${<}10\%$ & ${<}5\%$ & ${<}15\%$ \\ 
$\abs{\eta}$ selection & ${<}0.95$ & ${<}0.9$ & ${<}1$ \\ 
$N_{\text{pixel-hit}}$ selection & ${>}1$ & ${>}2$ & ${>}0$ \\ 
$\chi ^2 N^{-1}_{\text{dof}}N^{-1}_{\text{layer}}$ selection & ${<}0.18$ & ${<}0.15$ & ${<}0.18$ \\ 
$\abs{d_{\mathrm{xy}}/\sigma (d_{\mathrm{xy}})}$ selection & ${<}3$ & ${<}2$ & ${<}5$ \\ 
$\abs{d_{\mathrm{z}}/\sigma (d_{\mathrm{z}})}$ selection & ${<}3$ & ${<}2$ & ${<}5$ \\ 
$(\Delta \eta ,\Delta \phi)$ pair selection & \begin{tabular}[c]{@{}c@{}}$\Delta \eta _{\text{min}}=0.014$\\ $\Delta \phi _{\text{min}}=0.022$\end{tabular} & \begin{tabular}[c]{@{}c@{}}$\Delta \eta _{\text{min}}=0.017$\\ $\Delta \phi _{\text{min}}=0.028$\end{tabular} & \begin{tabular}[c]{@{}c@{}}$\Delta \eta _{\text{min}}=0.011$\\ $\Delta \phi _{\text{min}}=0.016$\end{tabular} \\ 
$q_{\text{min}}$ lower fit limit & $q_{\text{min}}^0$($\KT$, cent) & $q_{\text{min}}^0-0.004$ & $q_{\text{min}}^0+0.004$ \\ 
$q_{\text{max}}$ upper fit limit & $q_{\text{max}}^0$($\KT$, cent) & $0.85 q_{\text{max}}^0$ & $1.15 q_{\text{max}}^0$ \\ 
centrality edges & Default values & Lower values & Higher values \\ 
\end{scotch}
\label{t:syst}
\end{table*}

\begin{table*}[h!]
\centering
\topcaption{The relative effect of the different types of systematic sources in each centrality class for the $R$ (upper Table), $\alpha$ (middle Table) and $\lambda$ (lower Table) parameters (in percentage). The values were averaged over $\KT$ and the two charge signs; the upwards (downwards) arrow, $\uparrow$ ($\downarrow$) represents positive (negative) uncertainty in the value of the final fit parameters.}
\begin{scotch}{cccccccccc}
&  & \multicolumn{2}{c}{Track and event selection} & \multicolumn{2}{c}{Pair selection} & \multicolumn{2}{c}{Fit limits} & \multicolumn{2}{c}{Overall} \\ \hline
& Cent. [\%] & \multicolumn{1}{c}{$\uparrow$} & $\downarrow$ & \multicolumn{1}{c}{$\uparrow$} & $\downarrow$ & \multicolumn{1}{c}{$\uparrow$} & $\downarrow$ & \multicolumn{1}{c}{$\uparrow$} & $\downarrow$ \\ 
\multirow{6}{*}{$\delta R$ [\%]}& 0-5 & \multicolumn{1}{c}{1.0} & 0.4 & \multicolumn{1}{c}{0.0} & 2.9 & \multicolumn{1}{c}{2.5} & 3.2 & \multicolumn{1}{c}{2.7} & 4.3 \\ 
& 5--10 & \multicolumn{1}{c}{0.7} & 0.5 & \multicolumn{1}{c}{0.0} & 2.1 & \multicolumn{1}{c}{2.0} & 2.3 & \multicolumn{1}{c}{2.1} & 3.2 \\ 
& 10--20 & \multicolumn{1}{c}{0.6} & 0.4 & \multicolumn{1}{c}{0.0} & 1.4 & \multicolumn{1}{c}{1.8} & 2.3 & \multicolumn{1}{c}{1.9} & 2.7 \\ 
& 20--30 & \multicolumn{1}{c}{0.6} & 0.3 & \multicolumn{1}{c}{0.2} & 1.6 & \multicolumn{1}{c}{1.8} & 1.9 & \multicolumn{1}{c}{1.9} & 2.5 \\ 
& 30--40 & \multicolumn{1}{c}{0.5} & 0.2 & \multicolumn{1}{c}{0.6} & 1.5 & \multicolumn{1}{c}{1.9} & 1.7 & \multicolumn{1}{c}{2.1} & 2.3 \\ 
& 40--60 & \multicolumn{1}{c}{0.5} & 0.5 & \multicolumn{1}{c}{0.7} & 1.4 & \multicolumn{1}{c}{1.9} & 1.7 & \multicolumn{1}{c}{2.1} & 2.3 \\[\cmsTabSkip]
\multirow{6}{*}{$\delta \alpha$ [\%]}& 0-5 & \multicolumn{1}{c}{0.6} & 0.7 & \multicolumn{1}{c}{4.8} & 0.0 & \multicolumn{1}{c}{6.9} & 3.8 & \multicolumn{1}{c}{8.4} & 3.9 \\ 
& 5--10 & \multicolumn{1}{c}{1.1} & 0.6 & \multicolumn{1}{c}{3.2} & 0.0 & \multicolumn{1}{c}{5.2} & 3.0 & \multicolumn{1}{c}{6.2} & 3.1 \\ 
& 10--20 & \multicolumn{1}{c}{0.3} & 0.4 & \multicolumn{1}{c}{1.9} & 0.0 & \multicolumn{1}{c}{4.9} & 3.0 & \multicolumn{1}{c}{5.3} & 3.0 \\ 
& 20--30 & \multicolumn{1}{c}{0.1} & 0.5 & \multicolumn{1}{c}{2.0} & 0.1 & \multicolumn{1}{c}{5.1} & 2.8 & \multicolumn{1}{c}{5.5} & 2.8 \\ 
& 30--40 & \multicolumn{1}{c}{0.2} & 0.7 & \multicolumn{1}{c}{2.0} & 0.7 & \multicolumn{1}{c}{4.5} & 3.5 & \multicolumn{1}{c}{4.9} & 3.6 \\ 
& 40--60 & \multicolumn{1}{c}{0.4} & 0.4 & \multicolumn{1}{c}{1.7} & 0.7 & \multicolumn{1}{c}{4.4} & 4.0 & \multicolumn{1}{c}{4.7} & 4.1 \\[\cmsTabSkip]
\multirow{6}{*}{$\delta \lambda$ [\%]}& 0-5 & \multicolumn{1}{c}{4.7} & 2.0 & \multicolumn{1}{c}{0.0} & 6.8 & \multicolumn{1}{c}{6.1} & 8.9 & \multicolumn{1}{c}{7.7} & 11.4 \\ 
& 5--10 & \multicolumn{1}{c}{3.7} & 1.7 & \multicolumn{1}{c}{0.0} & 5.5 & \multicolumn{1}{c}{4.9} & 6.9 & \multicolumn{1}{c}{6.1} & 9.0 \\ 
& 10--20 & \multicolumn{1}{c}{2.6} & 1.3 & \multicolumn{1}{c}{0.0} & 4.3 & \multicolumn{1}{c}{4.7} & 6.5 & \multicolumn{1}{c}{5.4} & 7.9 \\ 
& 20--30 & \multicolumn{1}{c}{2.4} & 1.0 & \multicolumn{1}{c}{1.1} & 4.4 & \multicolumn{1}{c}{4.2} & 6.4 & \multicolumn{1}{c}{5.0} & 7.8 \\ 
& 30--40 & \multicolumn{1}{c}{2.4} & 0.9 & \multicolumn{1}{c}{1.7} & 4.0 & \multicolumn{1}{c}{4.9} & 5.6 & \multicolumn{1}{c}{5.7} & 6.9 \\ 
& 40--60 & \multicolumn{1}{c}{2.0} & 0.9 & \multicolumn{1}{c}{1.7} & 3.3 & \multicolumn{1}{c}{5.6} & 5.1 & \multicolumn{1}{c}{6.2} & 6.1 \\
\end{scotch}
\label{t:Rsyst}
\end{table*}

\section{Results} \label{sec:results}

Three physical parameters ($R$, $\alpha$, $\lambda$) were determined for each centrality and $\KT$ class 
by fitting the theoretical formula to the measured correlation functions. 
The results are presented as a function of \mT. Systematic uncertainties are separated 
into correlated and uncorrelated point-to-point parts. Correlated uncertainties associate the same relative uncertainty for a given parameter for all centrality and $\KT$ classes, 
whereas point-to-point values can vary between classes. This separation is necessary when investigating the \mT or centrality dependence of the fit parameters,
where only the uncorrelated point-to-point uncertainty is used in determining the $\chi^2$ values for the fits. 
The correlated part of the systematic uncertainty for each parameter value ($R$, $\lambda$, or $\alpha$) is taken as the average uncertainty for the parameter. The point-to-point uncertainty for a parameter value is then the difference between the full systematic and the correlated uncertainties.

\subsection{The L\'evy scale parameter \texorpdfstring{$R$}{R} }

\begin{figure*}
\centering
\includegraphics[width=0.9\textwidth]{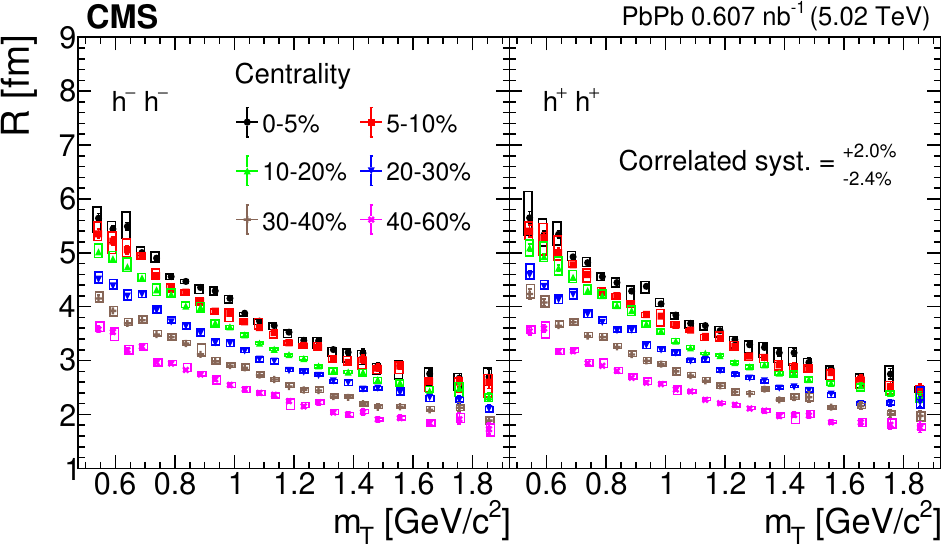}
\caption{The L\'evy scale parameter $R$ versus the transverse mass \mT in different centrality classes, for negatively (left) and positively (right) charged hadron pairs. The error bars show the statistical uncertainties, while the boxes indicate the point-to-point systematic uncertainties. These boxes are slightly shifted along the horizontal axes for better visibility. The correlated systematic uncertainty is also indicated.}
\label{f:R}
\end{figure*}

Figure~\ref{f:R} shows the obtained $R$ values as a function of \mT.
All values are found between 1.6 and 5.8\unit{fm}.
The decreasing trend in \mT that is expected from hydrodynamics is observed.  A clear centrality dependence is also visible. 
The decreasing value of $R$ as collisions become more peripheral supports the geometric interpretation of this scale parameter.

\begin{figure*}
\centering
\includegraphics[width=0.9\textwidth]{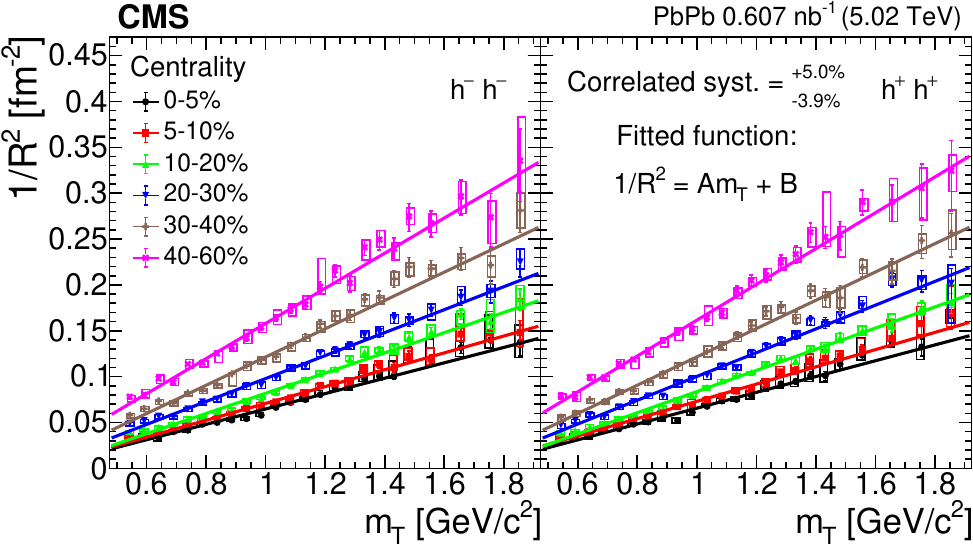}
\caption{The $1/R^2$ distribution \vs transverse mass \mT in different centrality classes, for negatively (left) and positively (right) charged hadron pairs. The error bars show the statistical uncertainties, while the boxes indicate the point-to-point systematic uncertainties. These boxes are slightly shifted along the horizontal axes for better visibility. The correlated systematic uncertainty is also indicated. A linear fit to the data is shown for each centrality bin. The fit parameters are tabulated in Table~\ref{t:1R2}.}
\label{f:1R2}
\end{figure*}

To further analyze the \mT dependence of $R$, $1/R^2$ was plotted versus \mT as shown in Fig.~\ref{f:1R2}. 
This plot is motivated  by hydrodynamic predictions~\cite{Csorgo:1995bi,Makhlin:1987gm} that suggest a linear dependence when a Gaussian source is assumed.
Here it is investigated whether a similar linear behavior exists for a L\'evy source. Linear fits are shown in Fig.~\ref{f:1R2}, with the fit parameters tabulated 
in Table~\ref{t:1R2}. Statistical and point-to-point systematic uncertainties were added in quadrature for the fits and used for the determination of the statistical uncertainties of the fit parameters. 
The confidence levels were statistically acceptable everywhere (\ie, have confidence level above 0.1\%, similarly to Refs.~\cite{PHENIX:2017ino, Achard:2011zza}), suggesting that hydrodynamic scaling of $1/R^2$ holds for a L\'evy source as well.

\begin{figure*}
\centering
\includegraphics[width=0.49\textwidth]{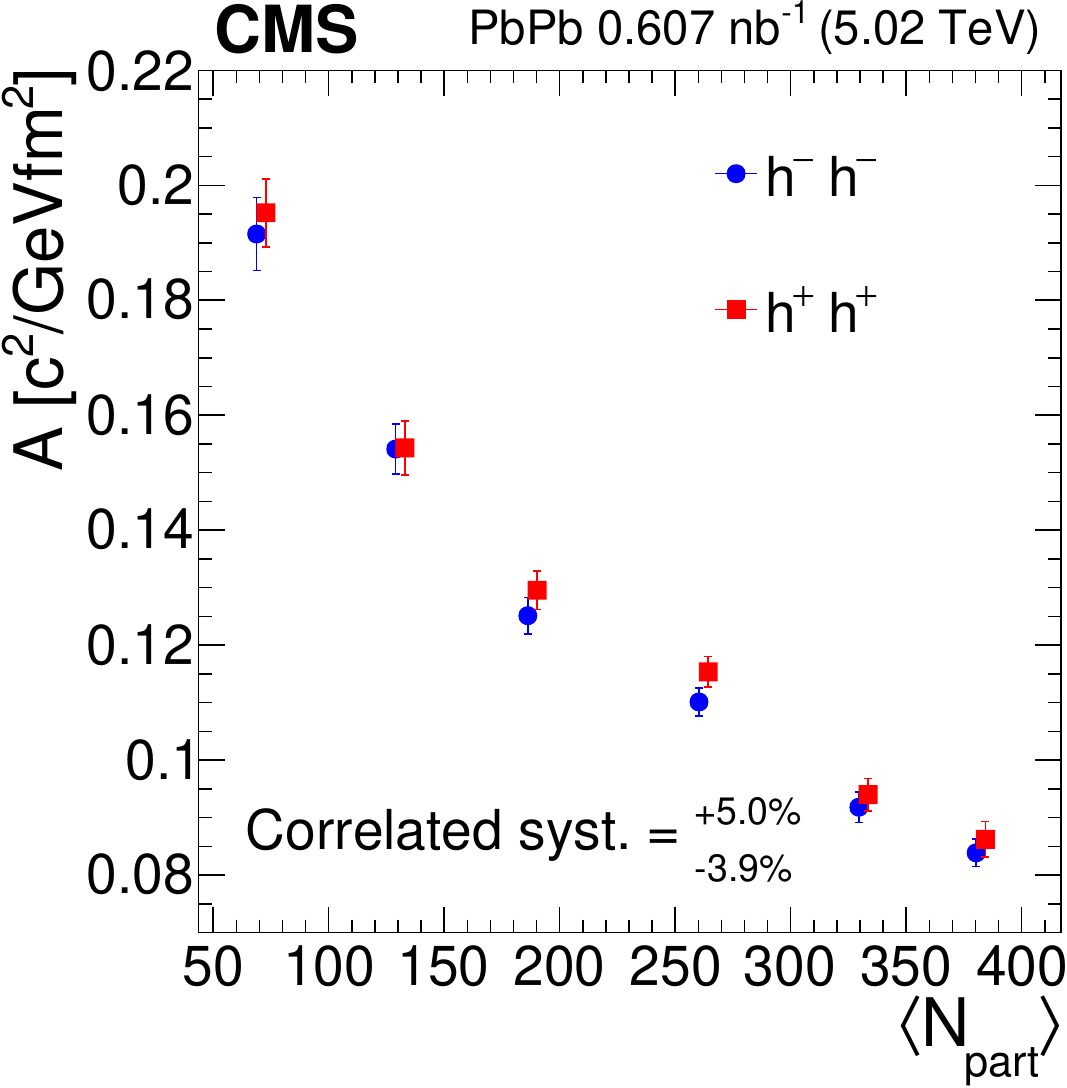} \hfill
\includegraphics[width=0.47\textwidth]{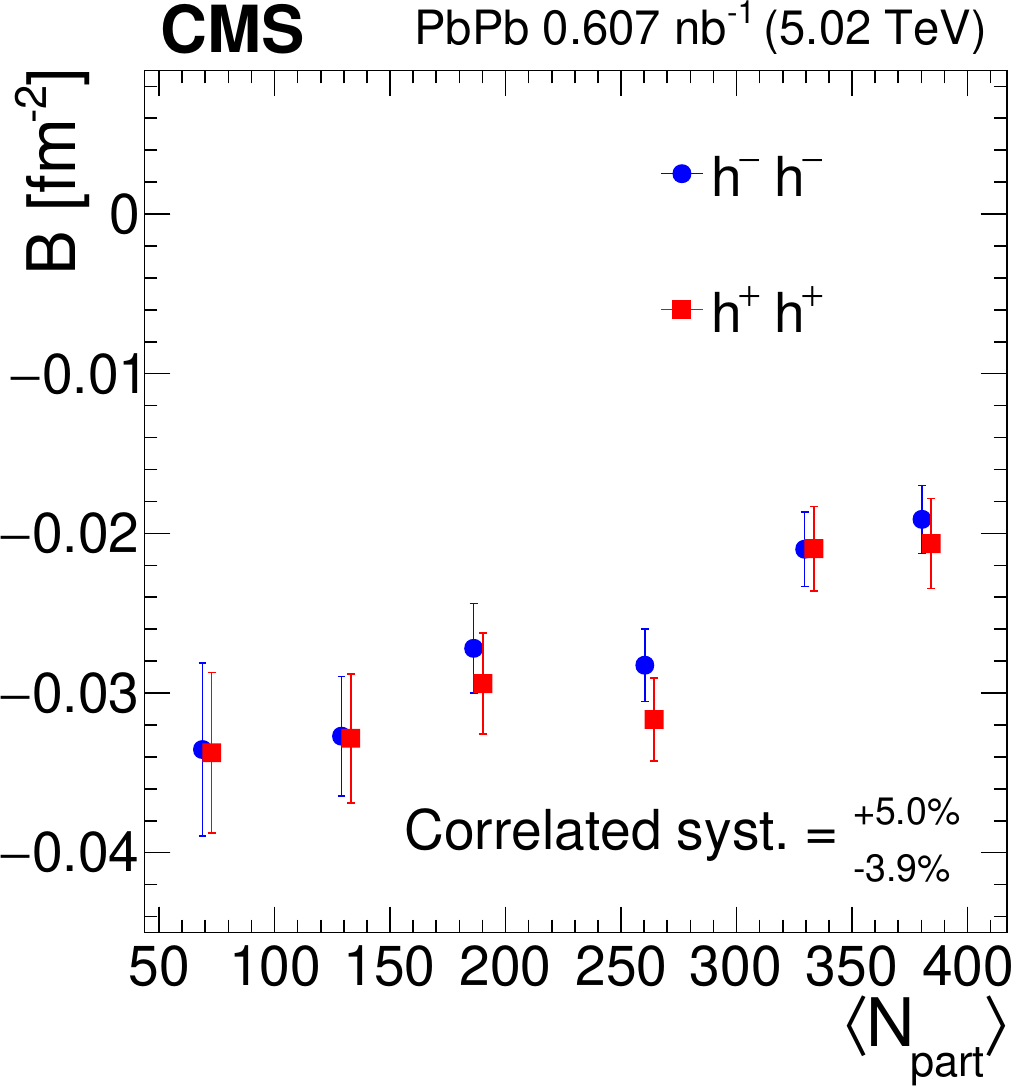}
\caption{The slope $A$ (left) and the intercept $B$ (right) linear-fit parameters versus \NpartAv, for positively and negatively charged hadron pairs. The error bars show the statistical uncertainties. The correlated systematic uncertainty is also indicated. The points are slightly shifted along the horizontal axes for better visibility. }
\label{f:ABfitparameters}
\end{figure*}

The slope $A$ and the intercept $B$ of the linear fits are also related to hydrodynamic models. 
The slope $A$ has connection to the Hubble constant ($H$) of the quark-gluon plasma~\cite{Csorgo:1995bi,Chojnacki:2004ec}, with
\begin{linenomath}
\begin{equation}
    A=\frac{H^2}{T_{\mathrm{f}}},
\end{equation}
\end{linenomath}
where $T_{\mathrm{f}}$ is the freeze-out temperature. With a Gaussian source assumption, the intercept $B$ is related to the size of the source at freeze-out ($R_{\mathrm{f}}$), with
\begin{linenomath}
\begin{equation}
    R_{\mathrm{f}}=\frac{1}{\sqrt{B}}.
\label{e:Rf}
\end{equation}
\end{linenomath}
Figure~\ref{f:ABfitparameters} shows that the slope parameter $A$ has a strong centrality dependence, since the average number of nucleons participating in the collision (\NpartAv) is closely related to centrality. The mapping from a centrality class to \NpartAv is given in Ref.~\cite{Loizides:2017ack}. 
By assuming a constant $T_{\mathrm{f}}\approx 156\MeV$~\cite{Steinbrecher:2018phh}, the value of the Hubble constant falls between 0.11--0.18\cOverfm, from the most central to the most peripheral collisions.
This indicates that the centrality of the collision affects the velocity of the expansion, resulting in larger velocities in more peripheral collisions. The values of the Hubble constant obtained here are close to the value of 0.17\cOverfm measured in high-multiplicity proton-proton collisions at $\sqrt{s}=13\TeV$~\cite{CMS:2019fur}, and are larger than the estimated value of 0.07\cOverfm measured in 0--30\% centrality AuAu collisions at $\sqrtsNN=200\GeV$~\cite{PHENIX:2017ino}.
The intercept parameter $B$ is found to be negative in all cases (as shown in Fig.~\ref{f:ABfitparameters}), which prevents a determination of the freeze-out size using Eq.~(\ref{e:Rf}). In the hydrodynamic models assuming a Gaussian shape, $B$ is positive. 
The negative value found here might be related to the L\'evy source assumed in the present analysis. Hydrodynamic solutions where $B$ is negative are possible~\cite{Csorgo:1998yk}, although 
the implications for a L\'evy source are not clear.

\begin{figure*}
\centering
\includegraphics[width=0.9\textwidth]{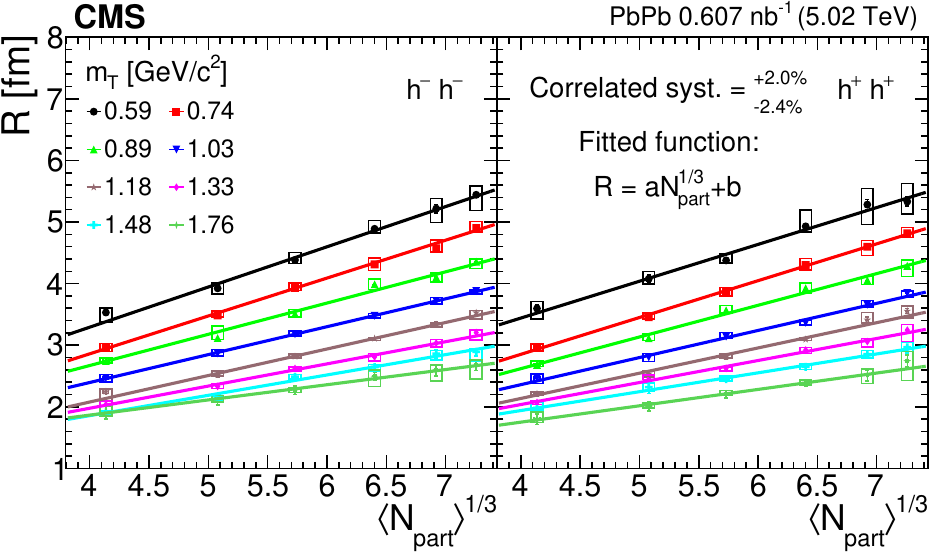}
\caption{The L\'evy scale parameter $R$ versus $\NpartAv^{1/3}$ in different \mT classes, for negatively (left) and positively (right) charged hadron pairs. The error bars show the statistical uncertainties, while the boxes indicate the point-to-point systematic uncertainties. The correlated systematic uncertainty is also indicated. A linear fit to the data is shown for each \mT class. The fit parameters are tabulated in Table~\ref{t:RNpart}.}
\label{f:RNpart}
\end{figure*}

Figure~\ref{f:RNpart} shows the dependence of $R$ on $\NpartAv^{1/3}$ for eight representative samples of the 24 \mT 
classes considered in this analysis, where the source volume is expected to scale with \NpartAv.  
The results are fitted with a linear function for each \mT class. The fits are shown in Fig.~\ref{f:RNpart}, 
with the fit parameters and the corresponding confidence levels tabulated 
in Table~\ref{t:RNpart}. The results are again consistent with a geometrical interpretation of the L\'evy scale parameter $R$.

\begin{table*}[h!]
\centering
\topcaption{Fit parameters and the corresponding confidence levels (\CL) of the linear fits to $1/R^2$ versus \mT, for positively (left) and negatively (right) charged hadron pairs.}
\cmsTable{
\begin{scotch}{ccccccc}
 & \multicolumn{3}{c}{$\mathrm{h}^+\mathrm{h}^+$}  & \multicolumn{3}{c}{$\mathrm{h}^-\mathrm{h}^-$} \\ \hline 
\multicolumn{1}{c}{Cent. [\%]} & \multicolumn{1}{c}{$A$ [$\unit{$c$}^2\unit{fm}^{-2}\GeV^{-1}$]} & \multicolumn{1}{c}{$B$ [${\unit{fm}}^{-2}$]} & \CL [\%] &  \multicolumn{1}{c}{$A$ [$\unit{$c$}^2\unit{fm}^{-2}\GeV^{-1}$]} & \multicolumn{1}{c}{$B$ [${\unit{fm}}^{-2}$]} & \CL [\%] \rule{0pt}{2.6ex} \rule[-1.2ex]{0pt}{0pt} \\ \hline
\multicolumn{1}{c}{0--5} & \multicolumn{1}{c}{$0.086\pm0.003$} & \multicolumn{1}{c}{$-0.021\pm0.003$} & 37.0  & \multicolumn{1}{c}{$0.084\pm0.002$} & \multicolumn{1}{c}{$-0.019\pm0.002$} & 10.5 \\ 
\multicolumn{1}{c}{5--10} & \multicolumn{1}{c}{$0.094\pm0.003$} & \multicolumn{1}{c}{$-0.021\pm0.003$} & 92.5  & \multicolumn{1}{c}{$0.092\pm0.003$} & \multicolumn{1}{c}{$-0.021\pm0.002$} & 78.1 \\ 
\multicolumn{1}{c}{10--20} & \multicolumn{1}{c}{$0.115\pm0.003$} & \multicolumn{1}{c}{$-0.032\pm0.003$} & 22.2  & \multicolumn{1}{c}{$0.110\pm0.002$} & \multicolumn{1}{c}{$-0.028\pm0.002$} & 1.0 \\ 
\multicolumn{1}{c}{20--30} & \multicolumn{1}{c}{$0.130\pm0.003$} & \multicolumn{1}{c}{$-0.029\pm0.003$} & 13.3  & \multicolumn{1}{c}{$0.125\pm0.003$} & \multicolumn{1}{c}{$-0.027\pm0.003$} & 23.2 \\ 
\multicolumn{1}{c}{30--40} & \multicolumn{1}{c}{$0.154\pm0.005$} & \multicolumn{1}{c}{$-0.033\pm0.004$} & 64.6  & \multicolumn{1}{c}{$0.154\pm0.004$} & \multicolumn{1}{c}{$-0.033\pm0.004$} & 6.2 \\ 
\multicolumn{1}{c}{40--60} & \multicolumn{1}{c}{$0.195\pm0.006$} & \multicolumn{1}{c}{$-0.034\pm0.005$} & 79.7  & \multicolumn{1}{c}{$0.192\pm0.006$} & \multicolumn{1}{c}{$-0.034\pm0.005$} & 82.7 \\ 
\end{scotch}
}
\label{t:1R2}
\end{table*}

\begin{table*}[h!]
\centering
\topcaption{Fit parameters and the corresponding confidence levels (\CL) of the linear fits to $R$ versus $\NpartAv^{1/3}$, for positively (left) and negatively (right) charged hadron pairs.}
\begin{scotch}{ccccccc}
 & \multicolumn{3}{c}{$\mathrm{h}^+\mathrm{h}^+$}  & \multicolumn{3}{c}{$\mathrm{h}^-\mathrm{h}^-$} \\ \hline
\multicolumn{1}{c}{\mT~[\GeVcc]} & \multicolumn{1}{c}{$a$ [fm]} & \multicolumn{1}{c}{$b$ [fm]} & \CL [\%]  & \multicolumn{1}{c}{$a$ [fm]} & \multicolumn{1}{c}{$b$ [fm]} & \CL [\%] \\ \hline
\multicolumn{1}{c}{0.59} & \multicolumn{1}{c}{$0.60\pm0.08$} & \multicolumn{1}{c}{$1.05\pm0.44$} & 88.1  & \multicolumn{1}{c}{$0.65\pm0.06$} & \multicolumn{1}{c}{$0.68\pm0.35$} & 79.5 \\ 
\multicolumn{1}{c}{0.74} & \multicolumn{1}{c}{$0.60\pm0.03$} & \multicolumn{1}{c}{$0.46\pm0.15$} & 96.1  & \multicolumn{1}{c}{$0.62\pm0.03$} & \multicolumn{1}{c}{$0.40\pm0.17$} & 90.7 \\ 
\multicolumn{1}{c}{0.89} & \multicolumn{1}{c}{$0.51\pm0.03$} & \multicolumn{1}{c}{$0.56\pm0.18$} & 56.0  & \multicolumn{1}{c}{$0.51\pm0.02$} & \multicolumn{1}{c}{$0.64\pm0.12$} & 70.8 \\ 
\multicolumn{1}{c}{1.03} & \multicolumn{1}{c}{$0.44\pm0.03$} & \multicolumn{1}{c}{$0.60\pm0.18$} & 69.0  & \multicolumn{1}{c}{$0.45\pm0.02$} & \multicolumn{1}{c}{$0.58\pm0.15$} & 99.8 \\ 
\multicolumn{1}{c}{1.18} & \multicolumn{1}{c}{$0.41\pm0.02$} & \multicolumn{1}{c}{$0.49\pm0.11$} & 54.5  & \multicolumn{1}{c}{$0.43\pm0.04$} & \multicolumn{1}{c}{$0.35\pm0.23$} & 94.0 \\ 
\multicolumn{1}{c}{1.33} & \multicolumn{1}{c}{$0.36\pm0.03$} & \multicolumn{1}{c}{$0.61\pm0.16$} & 84.0  & \multicolumn{1}{c}{$0.36\pm0.03$} & \multicolumn{1}{c}{$0.54\pm0.14$} & 92.0 \\ 
\multicolumn{1}{c}{1.48} & \multicolumn{1}{c}{$0.31\pm0.03$} & \multicolumn{1}{c}{$0.71\pm0.18$} & 96.3  & \multicolumn{1}{c}{$0.33\pm0.03$} & \multicolumn{1}{c}{$0.53\pm0.16$} & 64.2 \\ 
\multicolumn{1}{c}{1.76} & \multicolumn{1}{c}{$0.27\pm0.04$} & \multicolumn{1}{c}{$0.69\pm0.26$} & 98.9  & \multicolumn{1}{c}{$0.25\pm0.06$} & \multicolumn{1}{c}{$0.88\pm0.32$} & 99.9 \\ 
\end{scotch}
\label{t:RNpart}
\end{table*}

\subsection{The L\'evy stability index \texorpdfstring{$\alpha$}{alpha}} \label{ss:alpha}

\begin{figure*}[htb!]
\centering
\includegraphics[width=0.9\textwidth]{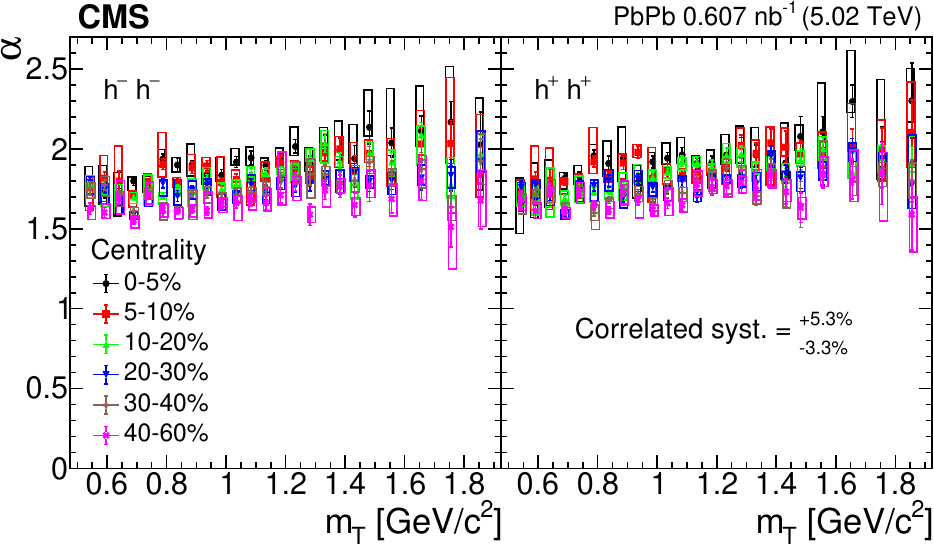}
\caption{The L\'evy stability index $\alpha$ versus the transverse mass \mT in different centrality classes, for negatively (left) and positively (right) charged hadron pairs. The error bars show the statistical uncertainties, while the boxes indicate the point-to-point systematic uncertainties. These boxes are slightly shifted along the horizontal axes for better visibility. The correlated systematic uncertainty is also indicated.}
\label{f:alpha}
\end{figure*}

\begin{figure}[htb!]
\centering
\includegraphics[width=0.5\textwidth]{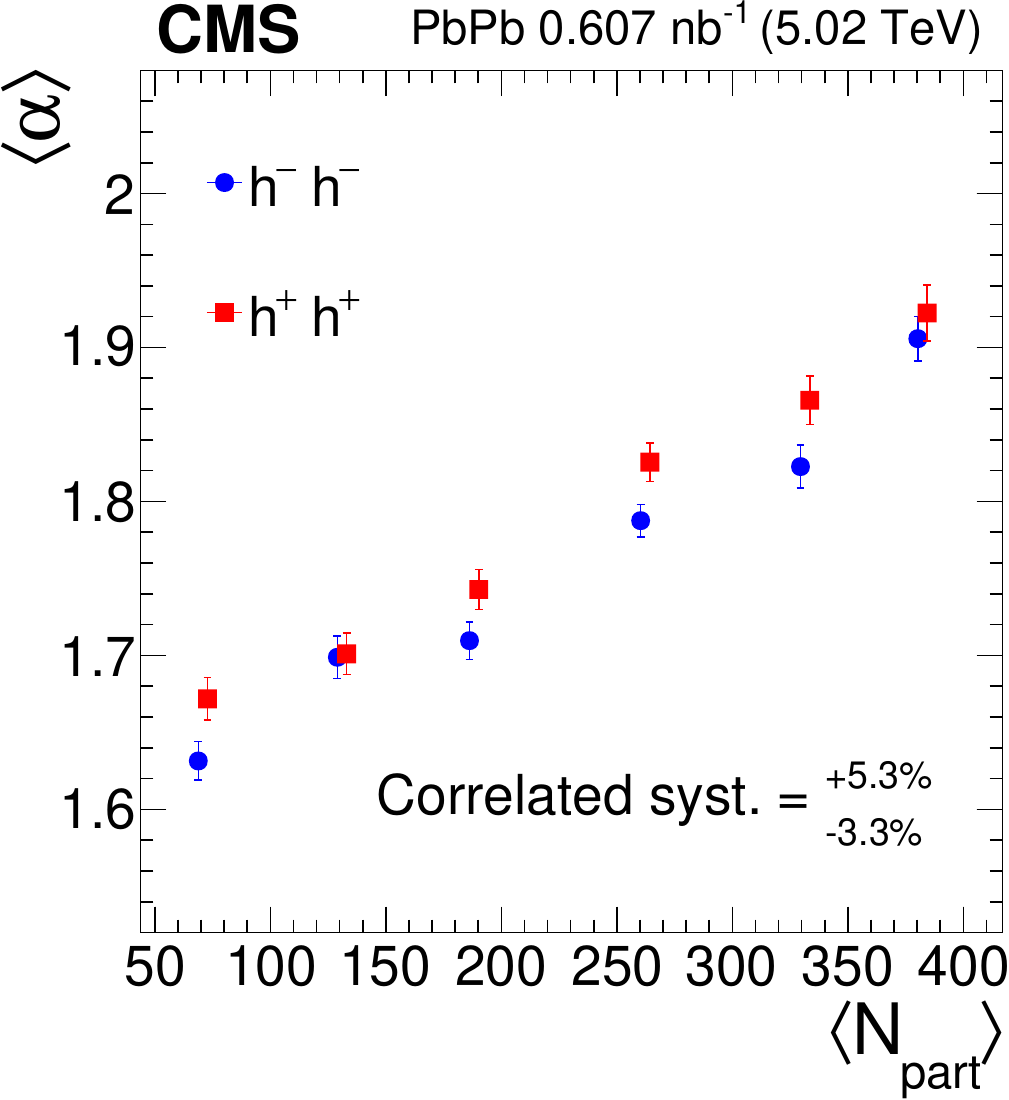}
\caption{The average L\'evy stability index \alphaAv versus \NpartAv, for both positively and negatively charged hadron pairs. The error bars show the statistical uncertainties. The correlated systematic uncertainty is also indicated. The points are slightly shifted along the horizontal axes for better visibility. }
\label{f:alphavsNpart}
\end{figure}

Figure~\ref{f:alpha} shows the values of $\alpha$ as a function of \mT for both negatively and positively charged hadron pairs. Within uncertainties, all values are found to fall in the range of 1.6--2.0, with little \mT dependence within a given centrality class. However, there is a clear centrality dependence, with the results tending to a Gaussian shape ($\alpha = 2$) for the most central events. The observation that the correlation functions for some centralities can be described with L\'evy functions having indices statistically inconsistent with a value of 2 means that a Gaussian assumption for the source shape is invalid in these cases. These L\'evy distribution fits can be interpreted as suggesting the presence of anomalous diffusion resulting from expansion in the hadron gas. To more clearly show the centrality dependence, as well as the significance of the deviation from 2.0, Fig.~\ref{f:alphavsNpart} shows \alphaAv (the brackets indicating an average over \mT) as a function of \NpartAv for both positively and negatively charged hadron pairs. A very similar linear dependence is observed for both charge signs.

The PHENIX Collaboration at RHIC reported a mean value for $\alpha$ of 1.207 for pions pairs with $\abs{\eta} < 0.35$ and $228 < \mT < 871\MeVcc$ in 0--30\% centrality AuAu collisions at $\sqrtsNN = 200\GeV$~\cite{PHENIX:2017ino}. Very little \mT dependence was observed at RHIC, a result similar to that reported here. However, our value for \alphaAv in the same centrality range is 45--60\% larger. This increase in the L\'evy stability index may be connected to the larger energy densities achieved at the LHC, which would be expected to result in less anomalous diffusion. An increasing energy density could also explain the increase of \alphaAv with \NpartAv.

\subsection{The correlation strength \texorpdfstring{$\lambda$}{lambda}} \label{ss:lambda}

\begin{figure*}[htb!]
\centering
\includegraphics[width=0.9\textwidth]{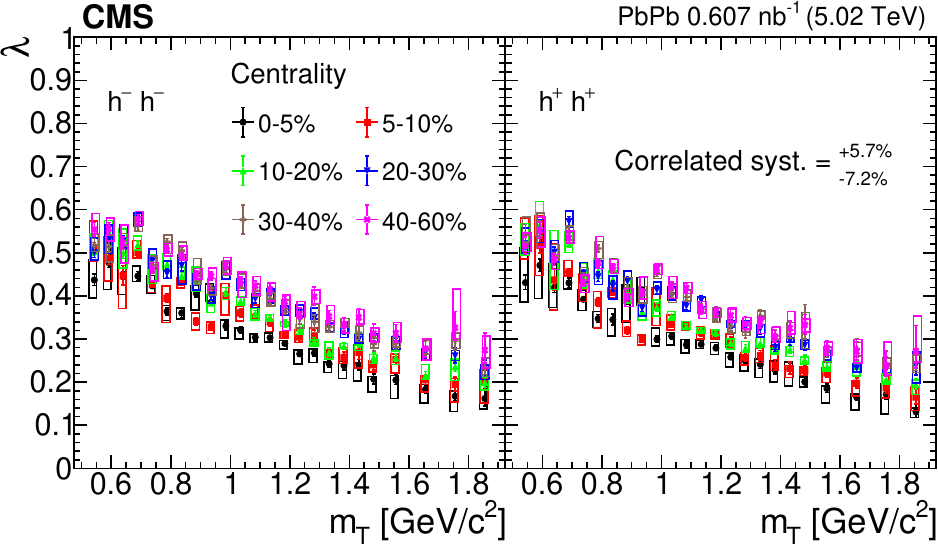}
\includegraphics[width=0.91\textwidth]{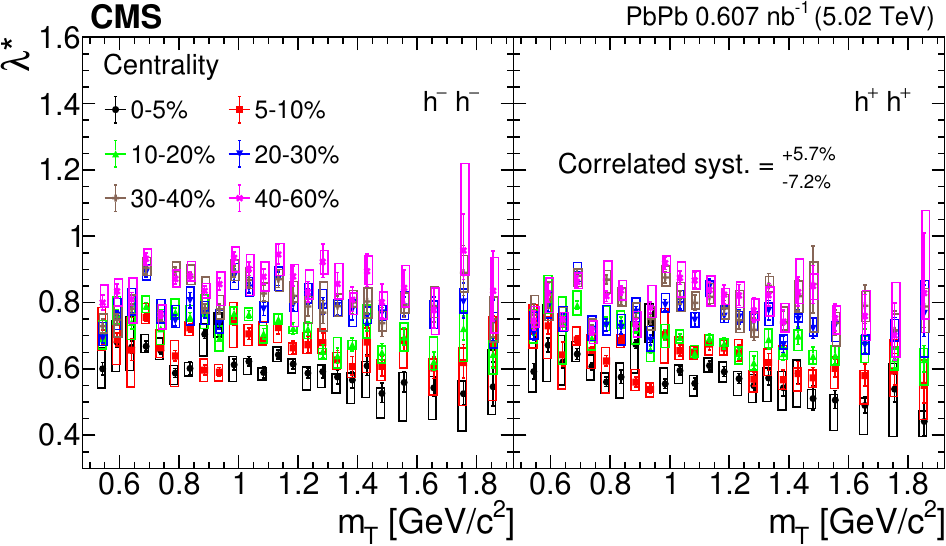}
\caption{The correlation strength $\lambda$ (upper panel) and $\lambda^*$, which is re-scaled with the square of the pion fraction (lower panel), versus the transverse mass \mT in different centrality classes, for negatively (left) and positively (right) charged hadron pairs. The error bars show the statistical uncertainties, while the boxes indicate the point-to-point systematic uncertainties. These boxes are slightly shifted along the horizontal axes for better visibility. The correlated systematic uncertainty is also indicated.}
\label{f:lambda}
\end{figure*}

The measured $\lambda$ values are shown in Fig.~\ref{f:lambda} as a function of \mT, for both negatively and positively charged hadron pairs. The values decrease with increasing transverse mass and as the collisions become more central.

The value of $\lambda$ is reduced compared with the case of using identified pions only, because the sample contains particles other than pions (mostly kaons and protons).
The strength of the correlation is proportional to the number of identical pion pairs, 
and pairs of nonidentical particles decrease the observed correlation. 
To see how the lack of particle identification affects the value of $\lambda$, 
a new parameter, $\lambda^*$, is introduced by re-scaling $\lambda$ with the square of the pion fraction:
\begin{linenomath}
\begin{equation}
    \lambda^*=\frac{\lambda}{(N_{\text{pion}}/N_{\text{hadron}})^2}.
\end{equation}
\end{linenomath}
The centrality and $\KT$ dependent pion fraction was measured by the ALICE Collaboration in the range $\abs{\eta}<0.5$, as reported in Ref.~\cite{ALICE:2019hno}.  The calculated $\lambda^*$ values as a function of \mT are shown in Fig.~\ref{f:lambda}. In each of the centrality classes and \mT bins, $\lambda^*$ is smaller than 1, which indicates that there exist additional effects lowering the correlation strength beyond the lack of particle identification discussed above. This result of $\lambda^*<1$ can, in fact, be understood on the basis of the core-halo model. In this interpretation, $\lambda^*$ describes the square of the fraction of pions produced in the core, as detailed in Eqs.~(\ref{e:C2:corehalo}) and (\ref{e:lambdadef}). Furthermore, $\lambda^*$ is found to lack any clear \mT dependence for a given centrality class. This is consistent with the observations at RHIC~\cite{PHENIX:2017ino}, and can be explained in terms of the core-halo model by the negligible momentum dependence of the core fraction in the investigated range. The decrease in $\lambda^*$ with centrality suggests a relatively smaller halo contribution for peripheral collisions.

\section{Summary} \label{sec:summary}

Two-particle Bose--Einstein momentum correlation function measurements are presented. The data sample consists of 4.27\ten{9} minimum bias
lead-lead (\PbPb) events, corresponding to an integrated luminosity of 0.607\nbinv, at a center-of-mass energy per nucleon pair of $\sqrtsNN = 5.02\TeV$, recorded by the CMS experiment at the LHC. 
The correlation functions found in different centrality and average transverse momentum classes
are analyzed in terms of L\'evy sources, including Coulomb effects. 
The values of the L\'evy scale parameter $R$, the L\'evy stability index $\alpha$, and the correlation strength $\lambda$ are determined. 

A geometric interpretation of the $R$ parameter is suggested by its dependence on the average number of participating nucleons in the collision. 
Assuming a pion mass for the charged particles, a linear dependence of $1/R^2$ on the transverse mass (\mT) is observed, consistent with a hydrodynamic scaling behavior, even for the case of L\'evy sources. 
Based on the observed linear behavior,  it is  estimated that the 
Hubble constant of the quark-gluon plasma created in 5.02\TeV \PbPb collisions increases from 0.11\cOverfm to 0.18\cOverfm when moving from most central to most peripheral collisions.
The intercept of the $1/R^2$ versus \mT linear fits is negative in all cases, requiring further studies for its interpretation. 
The $\alpha$ parameter is found to have little, if any, \mT dependence and to range between 1.6--2.0, 
increasing with centrality. The $\alpha$ values found in this paper are approximately 45--60\% larger than those
reported for gold-gold collisions at $\sqrtsNN = 200\GeV$ at RHIC. This increase, while not fully understood, may result from the greater energy
densities achieved at the LHC. As a function of \mT, a strong and decreasing trend is observed for the $\lambda$ parameter, which can be explained by the lack of particle identification. 
After rescaling the $\lambda$ values to account for the fraction of pions among the charged hadrons, a nearly constant trend of the pion-fraction-corrected $\lambda^*$ with \mT is observed. The $\lambda^*$ values are found to be smaller than unity, which can be interpreted on the basis of the core-halo model as a nonnegligible halo contribution. Furthermore, $\lambda^*$ is found to decrease as the collisions become more central.
Altogether, these results imply that the hadron emitting source in $\sqrtsNN = 5.02\TeV$  
\PbPb collisions can be described by L\'evy distributions.  
This allows for a new and precise characterization of this source in high-energy heavy ion collisions.

\begin{acknowledgments}
We congratulate our colleagues in the CERN accelerator departments for the excellent performance of the LHC and thank the technical and administrative staffs at CERN and at other CMS institutes for their contributions to the success of the CMS effort. In addition, we gratefully acknowledge the computing  centers and personnel of the Worldwide LHC Computing Grid and other  centers for delivering so effectively the computing infrastructure essential to our analyses. Finally, we acknowledge the enduring support for the construction and operation of the LHC, the CMS detector, and the supporting computing infrastructure provided by the following funding agencies: BMBWF and FWF (Austria); FNRS and FWO (Belgium); CNPq, CAPES, FAPERJ, FAPERGS, and FAPESP (Brazil); MES and BNSF (Bulgaria); CERN; CAS, MoST, and NSFC (China); MINCIENCIAS (Colombia); MSES and CSF (Croatia); RIF (Cyprus); SENESCYT (Ecuador); MoER, ERC PUT and ERDF (Estonia); Academy of Finland, MEC, and HIP (Finland); CEA and CNRS/IN2P3 (France); BMBF, DFG, and HGF (Germany); GSRI (Greece); NKFIH (Hungary); DAE and DST (India); IPM (Iran); SFI (Ireland); INFN (Italy); MSIP and NRF (Republic of Korea); MES (Latvia); LAS (Lithuania); MOE and UM (Malaysia); BUAP, CINVESTAV, CONACYT, LNS, SEP, and UASLP-FAI (Mexico); MOS (Montenegro); MBIE (New Zealand); PAEC (Pakistan); MES and NSC (Poland); FCT (Portugal); MESTD (Serbia); MCIN/AEI and PCTI (Spain); MOSTR (Sri Lanka); Swiss Funding Agencies (Switzerland); MST (Taipei); MHESI and NSTDA (Thailand); TUBITAK and TENMAK (Turkey); NASU (Ukraine); STFC (United Kingdom); DOE and NSF (USA).

\hyphenation{Rachada-pisek} Individuals have received support from the Marie-Curie program and the European Research Council and Horizon 2020 Grant, contract Nos.\ 675440, 724704, 752730, 758316, 765710, 824093, 884104, and COST Action CA16108 (European Union); the Leventis Foundation; the Alfred P.\ Sloan Foundation; the Alexander von Humboldt Foundation; the Belgian Federal Science Policy Office; the Fonds pour la Formation \`a la Recherche dans l'Industrie et dans l'Agriculture (FRIA-Belgium); the Agentschap voor Innovatie door Wetenschap en Technologie (IWT-Belgium); the F.R.S.-FNRS and FWO (Belgium) under the ``Excellence of Science -- EOS" -- be.h project n.\ 30820817; the Beijing Municipal Science \& Technology Commission, No. Z191100007219010; the Ministry of Education, Youth and Sports (MEYS) of the Czech Republic; the Hellenic Foundation for Research and Innovation (HFRI), Project Number 2288 (Greece); the Deutsche Forschungsgemeinschaft (DFG), under Germany's Excellence Strategy -- EXC 2121 ``Quantum Universe" -- 390833306, and under project number 400140256 - GRK2497; the Hungarian Academy of Sciences, the New National Excellence Program - \'UNKP, the NKFIH research grants K 124845, K 124850, K 128713, K 128786, K 129058, K 131991, K 133046, K 138136, K 143460, K 143477, 2020-2.2.1-ED-2021-00181, and TKP2021-NKTA-64 (Hungary); the Council of Science and Industrial Research, India; the Latvian Council of Science; the Ministry of Education and Science, project no. 2022/WK/14, and the National Science Center, contracts Opus 2021/41/B/ST2/01369 and 2021/43/B/ST2/01552 (Poland); the Funda\c{c}\~ao para a Ci\^encia e a Tecnologia, grant CEECIND/01334/2018 (Portugal); the National Priorities Research Program by Qatar National Research Fund; MCIN/AEI/10.13039/501100011033, ERDF ``a way of making Europe", and the Programa Estatal de Fomento de la Investigaci{\'o}n Cient{\'i}fica y T{\'e}cnica de Excelencia Mar\'{\i}a de Maeztu, grant MDM-2017-0765 and Programa Severo Ochoa del Principado de Asturias (Spain); the Chulalongkorn Academic into Its 2nd Century Project Advancement Project, and the National Science, Research and Innovation Fund via the Program Management Unit for Human Resources \& Institutional Development, Research and Innovation, grant B05F650021 (Thailand); the Kavli Foundation; the Nvidia Corporation; the SuperMicro Corporation; the Welch Foundation, contract C-1845; and the Weston Havens Foundation (USA).
\end{acknowledgments}

\ifthenelse{\boolean{cms@external}}{\clearpage}{}

\bibliography{auto_generated}

\cleardoublepage \appendix\section{The CMS Collaboration \label{app:collab}}\begin{sloppypar}\hyphenpenalty=5000\widowpenalty=500\clubpenalty=5000
\cmsinstitute{Yerevan Physics Institute, Yerevan, Armenia}
{\tolerance=6000
A.~Tumasyan\cmsAuthorMark{1}\cmsorcid{0009-0000-0684-6742}
\par}
\cmsinstitute{Institut f\"{u}r Hochenergiephysik, Vienna, Austria}
{\tolerance=6000
W.~Adam\cmsorcid{0000-0001-9099-4341}, J.W.~Andrejkovic, T.~Bergauer\cmsorcid{0000-0002-5786-0293}, S.~Chatterjee\cmsorcid{0000-0003-2660-0349}, K.~Damanakis\cmsorcid{0000-0001-5389-2872}, M.~Dragicevic\cmsorcid{0000-0003-1967-6783}, A.~Escalante~Del~Valle\cmsorcid{0000-0002-9702-6359}, P.S.~Hussain\cmsorcid{0000-0002-4825-5278}, M.~Jeitler\cmsAuthorMark{2}\cmsorcid{0000-0002-5141-9560}, N.~Krammer\cmsorcid{0000-0002-0548-0985}, L.~Lechner\cmsorcid{0000-0002-3065-1141}, D.~Liko\cmsorcid{0000-0002-3380-473X}, I.~Mikulec\cmsorcid{0000-0003-0385-2746}, P.~Paulitsch, J.~Schieck\cmsAuthorMark{2}\cmsorcid{0000-0002-1058-8093}, R.~Sch\"{o}fbeck\cmsorcid{0000-0002-2332-8784}, D.~Schwarz\cmsorcid{0000-0002-3821-7331}, M.~Sonawane\cmsorcid{0000-0003-0510-7010}, S.~Templ\cmsorcid{0000-0003-3137-5692}, W.~Waltenberger\cmsorcid{0000-0002-6215-7228}, C.-E.~Wulz\cmsAuthorMark{2}\cmsorcid{0000-0001-9226-5812}
% ...
\par}
\vskip\cmsinstskip
\dag:~Deceased\\
$^{1}$Also at Yerevan State University, Yerevan, Armenia\\
$^{2}$Also at TU Wien, Vienna, Austria\\
% ...
\end{sloppypar}
\end{document}